\documentclass[prl,twocolumn,showpacs]{revtex4}

\usepackage{graphicx}

\newcommand{\vecbf}[1]{\mbox{\boldmath{$#1$}}}

\begin{document}

\title{Spin-density induced by
electromagnetic wave in two-dimensional electron gas with both
Rashba and Dresselhaus spin-orbit couplings}

\author{Mikhail Pletyukhov}
\affiliation{Institut f\"ur Theoretische Physik A, Physikzentrum,
              RWTH Aachen, D-52056 Aachen, Germany}
\author{Alexander Shnirman}
\affiliation{Institut f\"ur Theorie der Kondensierten Materie and 
DFG Center for Functional Nanostructures (CFN),
Universit\"at Karlsruhe, D-76128 Karlsruhe, Germany}

\begin{abstract}
We consider the magnetic response of a two-dimensional
electron gas (2DEG) with both Rashba and Dresselhaus spin-orbit coupling 
to a microwave excitation. We generalize the results of Ref.~\cite{ShnirmanMartin}, 
where pure Rashba coupling was studied. 
We observe that the microwave with the in-plane electric field and the out-of-plane 
magnetic field creates an out-of-plane spin polarization. 
The effect is more prominent in clean systems with resolved 
spin-orbit-split subbands. Considered as response to the microwave 
magnetic field, the spin-orbit contribution to the magnetization 
far exceeds the usual Zeeman contribution in the clean limit. The effect vanishes when the 
Rashba and the Dresselhaus couplings have equal strength.
\end{abstract}

\pacs{72.25.Rb,85.75.-d}

\maketitle

The spin-orbit effects in semiconductors have been studied for a
long time~\cite{DyakonovPerel,aronov_lyanda,Edelstein}.
The discussion was revived in relation to the spin-Hall 
effect~\cite{Sinova,Murakami} in hope of applying spin-orbit related effects to
spintronics~\cite{murakami-2005-45}. Initially the effect was considered for conductors,
e.g., for a 2DEG. Although the attention of the 
community has now mostly switched to the quantum spin-Hall effect in 
"spin-Hall insulators"~\cite{PhysRevLett.93.156804,kane:146802}, we consider 
here the spin-Hall related effects in a 2DEG with both Rashba and Dresselhaus 
spin-orbit couplings.

The promise of the spin-Hall effect is in the possibility of generation of 
non-equilibrium spin polarization by means of a DC electric field.  
However, after some discussions it was concluded that in two-dimensional electron
gases with the Rashba and the Dresselhaus spin-orbit interactions
the spin-Hall effect vanishes for constant and homogeneous
electric field~\cite{Inoue,Mishchenko,Dimitrova,chalaev-2005-71}.
At finite frequencies the spin-Hall effect is non-zero~\cite{Mishchenko}.

While for the homogeneous spin-Hall effect (uniform applied electric field) the
out-of-plane spin polarization is expected to accumulate only at the edges of
the sample~\cite{Mishchenko}, there is an alternative possibility that we
explore here: To create an out-of-plane inhomogeneous spin density
{\em in the bulk} in response to a spatially-modulated field.  Such bulk
accumulation is free from the uncertainties associated with the charge and spin
transport near the sample boundaries, and thus may provide an unambiguous
method to detect the spin-Hall effect.

Alternatively, one can consider our results as providing the spin response 
to a long wave (spatially homogeneous), out-of-plane, oscillating magnetic field. 
According to the Faraday induction law such a field creates spatially inhomogeneous, 
in-plane electric field which in turn is responsible for the effect.     

In this paper we generalize the results of Ref.~\cite{ShnirmanMartin} where the
magnetic response of a 2DEG with pure Rashba spin-orbit coupling to microwaves 
was studied. Here we consider a more general situation where both couplings are present.
 
The spin-orbit coupling reads
\begin{equation}
H_{\rm SO}=\alpha_{\rm R}(-p_x \sigma_y + p_y \sigma_x) +
\alpha_{\rm D}(p_x \sigma_x - p_y \sigma_y) \ ,
\end{equation}
where $\alpha_{\rm R}$ and $\alpha_{\rm D}$ are the strengths of the 
Rashba and Dresselhaus spin-orbit couplings respectively.
It is convenient to perform a $\pi/4$ rotation in both the momentum 
and the spin spaces. That is 
$
p'_x = \frac{p_x+p_y}{\sqrt{2}} \quad , \quad
p'_y =  \frac{-p_x+p_y}{\sqrt{2}}
$, 
and
$
\sigma'_x = \frac{\sigma_x+\sigma_y}{\sqrt{2}}\quad , \quad
\sigma'_y = \frac{-\sigma_x+\sigma_y}{\sqrt{2}}
$.
Then the spin-orbit coupling reads
\begin{equation}
H_{\rm SO}=-(\alpha_{\rm R}+\alpha_{\rm D}) p'_x \sigma'_y 
+(\alpha_{\rm R}-\alpha_{\rm D}) p'_y \sigma'_x \ .
\end{equation}
In what follows we work in the rotated basis and omit the primes.

Introducing the angle $\phi_p$ via $\vecbf{p}=|\vecbf{p}|(\cos\phi_p,\sin\phi_p)$
we obtain the energies of the two sub-bands given by
$\epsilon^{\pm}(\vecbf{p}) = \frac{\vecbf{p}^2}{2m^*}\pm
\alpha_\phi(\phi_p) |\vecbf p|$, 
where $m^*$ is the electron band mass, $\alpha_\phi(\phi_p) \equiv 
\sqrt{(\alpha_{\rm R}^2+\alpha_{\rm
D}^2)(1+T_\alpha\cos 2\phi_p)}$, and $T_\alpha\equiv \frac{2\alpha_{\rm
R}\alpha_{\rm D}}{\alpha_{\rm R}^2+\alpha_{\rm D}^2}$. For purely Rashba 
(Dresselhaus) coupling 
$T_\alpha = 0$, while in the case of equal coupling strengths $T_\alpha = 1$.

We consider a linearly polarized
in-plane microwave field $\vecbf A = \vecbf A_0 \exp(i \vecbf q
\vecbf r-i\Omega t)$, where $\vecbf A_0 = A_0 (\cos\alpha,\sin\alpha,0)$ and 
$\vecbf q = q(\sin\alpha,-\cos\alpha,0)$. The signs are chosen so that for positive 
$A_0$ and $q$ the 
vectors $\vecbf q,\vecbf A_0,\vecbf e_z$ form a right-handed basis. As usual
$\vecbf E = (i\Omega/c)\vecbf A$ and $\vecbf B = i \vecbf q\times \vecbf A$.

{\it Kinetic equation.} 
We follow the route of Refs.~\cite{Burkov,Mishchenko,ShnirmanMartin} and
use the standard linear response Keldysh technique to determine the dynamics 
of the charge and spin densities. 
For introduction into the Keldysh technique see Ref.~\cite{RammerSmith}. 
In the dirty limit (to be defined below) this technique leads to diffusion equations
for the spin and charge densities~\cite{Mishchenko,stanescu:125307}. We 
concentrate here on the clean limit although our results are valid also in the 
dirty limit.    

We consider only the s-wave disorder scattering, that is $V_{\rm
disorder} = \sum_k u\delta(\vecbf r - \vecbf r_k)$ where $\vecbf
r_k$ are random locations with the average density $n_{\rm imp}$.
We employ the linear response,  $H=H_0+H_1$, with 
\begin{equation}
H_0= \frac{\vecbf p^2}{2m^*} + \vecbf\eta \vecbf p  + V_{\rm disorder}\ ,
\end{equation}
where $\vecbf \eta = \left[(-(\alpha_{\rm R}+\alpha_{\rm D})\sigma_y,
(\alpha_{\rm R}-\alpha_{\rm D})\sigma_x\right]$, 
and
\begin{eqnarray}
H_1 &=& -\frac{e}{2c}\,\left\{\vecbf v , \vecbf A\right\}_+ 
-\frac{1}{2}g\mu_{\rm B} \vecbf B \vecbf \sigma \nonumber\ ,
\end{eqnarray}
where $\vecbf v \equiv \frac{\vecbf p}{m^*}+\vecbf\eta$.

The zeroth order in $\vecbf A$ Green's functions, $G_0$,
reflect the standard disorder broadening. We introduce the inverse momentum relaxation time
$\tau^{-1}=2\pi n_{\rm imp} u^2 \nu$, where $n_{\rm imp}$ is the density of impurities 
and $\nu = m^*/(2\pi\hbar^2)$ is the electronic density of states per spin 
(strictly speaking $\nu$ is the density of states in absence of the spin-orbit coupling).
We obtain
\begin{equation}
G_0^R = \left(\frac{1}{2}+\frac{1}{2}\frac{\vecbf\eta \vecbf
p}{\alpha_\phi |\vecbf p|}\right)G_0^{R+} +
\left(\frac{1}{2}-\frac{1}{2}\frac{\vecbf\eta \vecbf
p}{\alpha_\phi |\vecbf p|}\right)G_0^{R-}\ ,
\end{equation}
where $G_0^{R\pm}(\vecbf p,\omega)\equiv \left(\omega -
\epsilon^{\pm}(\vecbf p)+i/(2\tau)\right)^{-1}$. In equilibrium
$G^K_0 = h(\omega)\,(G^R_0-G^A_0)$, where $h(\omega) \equiv
\tanh\frac{\omega-E_F}{2T}$.

Within the self-consistent Born approximation we find the 
linear (in $\vecbf A$) correction to the Green's function, 
$G_1$. Having $G_1$ we can calculate any single-particle quantity, i.e.,
density or current. As usual in the linear response theory the Keldysh component
$G^K_1$ splits into two parts, $G^K_1=G^{K,I}_1+G^{K,II}_1$.
The first part, $G^{K,I}_1$, corresponds to the retarded-advanced (R-A) 
combinations in the Kubo
formula, while $G^{K,II}_1$ stands for the R-R and A-A
combinations~\cite{RammerSmith}.
The standard Keldysh perturbation theory gives
the spin-charge density matrix as $\hat \rho = \frac{1}{2}\,n(\vecbf q,\Omega) + \vecbf s(\vecbf q,\Omega) \vecbf \sigma= \int
\frac{d\omega}{2\pi} \int \frac{d^2 p}{(2\pi)^2} \,[-i\, G_1^<] =
-\frac{i}{2} \int \frac{d\omega}{2\pi} \int \frac{d^2
p}{(2\pi)^2}(G_1^K - G_1^R + G_1^A)$, where $n(\vecbf q,\Omega)$ is 
the charge density while
$\vecbf s(\vecbf q,\Omega)$ is the spin density. It splits as follows  
$\hat\rho=\hat\rho^{I}+\hat\rho^{II}$,
where $\hat\rho^{I}=-\frac{i}{2} \int \frac{d\omega}{2\pi} \int
\frac{d^2 p}{(2\pi)^2}\,G_1^{K,I}$ and $\hat\rho^{II} =-\frac{i}{2} \int \frac{d\omega}{2\pi} \int \frac{d^2
p}{(2\pi)^2}(G_1^{K,II} - G_1^R + G_1^A)$. 
 
Introducing the average spin-orbit band splitting $\Delta_F \equiv p_F\sqrt{\alpha_{\rm R}^2+\alpha_{\rm D}^2}$, we can define three regimes:  (i) 
``super-clean'' $\tau^{-1}< \Delta_F^2 m^*/p_F^2=m^*(\alpha_{\rm R}^2+\alpha_{\rm D}^2)$; 
(ii) clean $m^*(\alpha_{\rm R}^2+\alpha_{\rm D}^2)<\tau^{-1}<\Delta_F$;
(iii) dirty $\tau^{-1}>\Delta_F$.
Our results below apply both in the clean and in the dirty regimes, but not in the 
"super-clean" one, i.e., our results apply for  $\tau^{-1} > \Delta_F^2 m^*/p_F^2$.

We obtain
\begin{equation} \label{Eq:kin_rho}
\frac{(1-I)}{\tau}\hat\rho^{I} = i\Omega \nu\tilde
I\left[\frac{e\{\vecbf v \vecbf A\}_+}{2c}+\frac{1}{2}g\mu_{\rm B} \vecbf B \vecbf \sigma \right]\ ,
\end{equation}
The functional $\tilde I$ is defined as
\begin{eqnarray}
&&\tilde I[X(\vecbf p)]=\nonumber\\&&= \frac{1}{m^*\tau}\int
\frac{d^2p}{(2\pi)^2}\, G^R_0(\vecbf p + \vecbf q/2,E_F
+\Omega/2)\cdot  X(\vecbf p) \cdot\nonumber\\&&\cdot\,
G^A_0(\vecbf p - \vecbf q/2,E_F -\Omega/2)\ ,
\end{eqnarray}
while $I$ is a $4\times 4$ matrix defined by its action on the
$4$-vector $\hat\rho$ as $I \hat\rho = \tilde I[\hat\rho]$ (we
just use the fact that $\hat\rho$ is independent of $\vecbf p$ to
represent the functional $\tilde I[\hat\rho]$ as a product of a
$4\times 4$ matrix $I$ and a vector $\hat\rho$).

The second contribution to the density, $\hat\rho^{II}$ is given
by
$
\hat\rho^{II}= \frac{1}{2} g\nu \mu_{\rm B} \vecbf B \vecbf \sigma
$.
Thus the total density follows from
\begin{eqnarray} \label{Eq:kin_rho_total_RHS}
\frac{(1-I)}{\tau}\hat\rho &=& i\Omega \nu\tilde
I\left[\frac{e\{\vecbf v \vecbf A\}_+}{2c}\right]\nonumber\\ &+&
\frac{[1-(1-i\Omega\tau) I]}{2\tau}\, g\nu \mu_{\rm B} \vecbf B \vecbf \sigma\ .
\end{eqnarray}

We allow for arbitrary external frequency $\Omega$, including
$\Omega> \tau^{-1}$.  However, we limit ourselves to the
experimentally relevant regime $v_F |\vecbf{q}| \ll \tau^{-1}$.
Recently the spin and charge response functions to the 
longitudinal fields were calculated for arbitrary values of
$\vecbf q$ in Ref.~\cite{Pletyukhov}.

We expand the matrix $I$ in powers of $\vecbf q$, $I = I^{(0)}+ I^{(1)}+\dots$.
In zeroth order in $\vecbf q$ the matrix $I$ is diagonal and 
its elements are given by
\begin{eqnarray}
I^{(0)}_{00} &=& \frac{1}{a}\quad,\quad I^{(0)}_{zz} = \frac{a}{\sqrt{R}}\ ,\nonumber\\
I^{(0)}_{xx} &=& \frac{(Q + D)-b^2(1+T_\alpha)}{2 a \sqrt{R}} \ ,\nonumber\\
I^{(0)}_{yy} &=& \frac{(Q - D)-b^2(1-T_\alpha)}{2 a \sqrt{R}} \ ,
\end{eqnarray}
where $R\equiv (a^2+b^2)^2 - b^4 T_\alpha^2 = (a^2+(1+T_\alpha)b^2)(a^2+(1-T_\alpha)b^2)$, $D\equiv \left(a^2+b^2 - \sqrt{R}\right)/T_\alpha$, 
and $Q\equiv \left(a^2+b^2 + \sqrt{R}\right)$. We have introduced $a\equiv
1-i\Omega\tau$ and $b\equiv 2\Delta_F\tau$.  Analyzing the path of the complex 
function $R(\Omega)$ we conclude that for $\sqrt{R}$ we have to choose 
the branch-cut along the positive semi-axis $R>0$. Alternatively we can use the usual 
definition of $\sqrt{\dots}$ (with the brunch-cut along the negative semi-axis) but replace $\sqrt{R}\rightarrow -i \sqrt{-R}$.

For the part linear in $\vecbf q$, $I^{(1)}$, we obtain the following matrix elements
\begin{eqnarray}
I^{(1)}_{zx}&=&-I^{(1)}_{xz} =\frac{p_F |\vecbf q|\tau \sin\alpha}{m^*}\, i^{(1)}_{zx} \ ,\nonumber\\
I^{(1)}_{zy}&=&-I^{(1)}_{yz}=\frac{p_F |\vecbf q| \tau \cos\alpha}{m^*}\, i^{(1)}_{zy}\ , 
\end{eqnarray}
where
\begin{eqnarray} 
i^{(1)}_{zx}&\equiv&  - \frac{i a b \sqrt{\frac{1+T_\alpha}{R}}}{a^2+b^2(1+T_\alpha)}
\ ,\nonumber\\
i^{(1)}_{zy}&\equiv&\phantom{-}\frac{i a b \sqrt{\frac{1-T_\alpha}{R}}}{a^2+b^2
   (1-T_\alpha)}\ .
\end{eqnarray}
We have neglected terms 
mixing the charge density with the spin density, since they contain a small 
parameter $m/(p_F^2\tau)$ as compared to the spin-spin terms.  

Expanding the RHS of
Eq.~(\ref{Eq:kin_rho_total_RHS}) in powers of $\vecbf q$ up to the
terms linear in $\vecbf q$ and neglecting again the charge density term 
for the same reason as above we obtain for the orbital term
\begin{eqnarray} \label{eq:C}
&&i\Omega \nu\tilde I\left[\frac{e\{\vecbf v \vecbf A\}_+}{2c}\right] =
C^{(0)}_{E,x}\sigma_x + C^{(0)}_{E,y}\sigma_y + C^{(1)}_{E,z}\sigma_z
\nonumber\\& 
=&\frac{\nu e \,|\vecbf E|  \sin\alpha}{\tau  p_F} c^{(0)}_{E,x}\sigma_x + 
\frac{\nu e \,|\vecbf E| \cos\alpha}{\tau  p_F} c^{(0)}_{E,y}\sigma_y\nonumber\\
&+& \frac{\nu e \,|\vecbf E| |\vecbf q| }{m^*} c^{(1)}_{E,z}(\alpha)\sigma_z
\ , 
\end{eqnarray}
where $\vecbf E =\frac{ i\Omega}{c} \vecbf A$.
The dimensionless coefficients in the 
zeroth order in $\vecbf q$ contributions are given by
\begin{eqnarray}
&&c^{(0)}_{E,x}=-
\frac{b(1+T_\alpha)\sqrt{\frac{1-T_\alpha}{R}}  
\left(b^2-D\right)}{4 a}\ , \nonumber\\
&&c^{(0)}_{E,y}=\phantom{-}
\frac{b (1-T_\alpha) \sqrt{\frac{1+T_\alpha}{R}} \left(b^2+D\right)}{4 a} \ ,
\end{eqnarray}
while the coefficient in the linear in $\vecbf q$ contribution is
\begin{eqnarray} 
&&c^{(1)}_{E,z}(\alpha)=i\,\sqrt{1-T_\alpha^2}\times\nonumber\\&&\times\frac{\left(b^4(1-T_\alpha^2)-a^4\right)  b^2-\left(2 T_\alpha a^2 b^4 +R D\right) \cos 2\alpha
}{4 a R \sqrt{R}}.\nonumber\\
\end{eqnarray}

Finally, the Zeeman term in the RHS of
Eq.~(\ref{Eq:kin_rho_total_RHS}) reads
\begin{eqnarray}
&&\frac{[1-(1-i\Omega\tau) I]}{2\tau}\, g\nu \mu_{\rm B} \vecbf B \vecbf \sigma
=\nonumber\\&&=
\frac{\nu\,g\,\mu_{\rm B}}{2\tau}
\sum_{\alpha=x,y,z} \left[1- a I^{(0)}_{\alpha\alpha} \right] B_\alpha\sigma_\alpha\ .
\end{eqnarray}
Note that the magnetic terms ($\propto \vecbf B=i\vecbf q\times \vecbf A$) are already 
of first order in $\vecbf q$.

We are now in a position to calculate the spin density. First we obtain the zeroth-order 
in $\vecbf q$ contribution. It is given by
\begin{eqnarray}
s_x^{(0)}&=&\frac{\tau}{1-I^{(0)}_{xx}} \, C^{(0)}_{E,x}\nonumber\\&=&-\frac{\nu e E_y}{2p_F}\cdot\frac{b (1+T_\alpha)\sqrt{1-T_\alpha}(b^2-D)}{2a\sqrt{R}-(Q + D)+b^2(1+T_\alpha)}\,, \nonumber \\
s_y^{(0)}&=&\frac{\tau}{1-I^{(0)}_{yy}} \, C^{(0)}_{E,y}\nonumber\\&=&\phantom{-}\frac{\nu e E_x}{2p_F}\cdot\frac{b(1-T_\alpha)\sqrt{1+T_\alpha}(b^2+D)}{2a\sqrt{R}-(Q - D)+b^2(1-T_\alpha)}\,, \nonumber \\
s_z^{(0)}&=&0\,.
\end{eqnarray}
This is a generalization of the well known
result~\cite{aronov_lyanda,Edelstein} meaning that there is an
in-plane spin polarization "perpendicular" to the applied electric field.
We observe that, for $T_\alpha \neq 0$, it is only perpendicular to $\vecbf E$ if the electric field
is along one of the main axes $x$ or $y$.

Next we calculate the first order orbital contribution. We only calculate the 
$z$ component of the spin density, as it was zero in the zeroth order. We obtain
\begin{eqnarray}\label{eq:s_z_orbital}
&&s_z^{(1),\rm orbital} = \frac{\tau}{1-I^{(0)}_{zz}} \, C^{(1)}_{E,z} \nonumber\\
&-& \frac{\tau}{1-I^{(0)}_{zz}}\,\left(-\frac{ I^{(1)}_{zx}}{\tau}\right)\,\frac{\tau}{1-I^{(0)}_{xx}}\,C^{(0)}_{E,x}
\nonumber\\
&-& \frac{\tau}{1-I^{(0)}_{zz}}\,\left(-\frac{ I^{(1)}_{zy}}{\tau}\right)\,\frac{\tau}{1-I^{(0)}_{yy}}\,C^{(0)}_{E,y}
\nonumber\\
&=& \frac{\tau C^{(1)}_{E,z} + I^{(1)}_{zx} s_x^{(0)}+I^{(1)}_{zy} s_y^{(0)}}{1-I^{(0)}_{zz}}\nonumber\\
&=&\frac{\nu e \,|\vecbf E| |\vecbf q| \tau}{m^*}\frac{1}{1-I^{(0)}_{zz}}\times\nonumber\\&\times&
\left[c^{(1)}_{E,z}(\alpha)
+\frac{ i^{(1)}_{zx}c_{E,x}^{(0)}\sin^2\alpha}{1-I^{(0)}_{xx}}
+\frac{ i^{(1)}_{zy}c_{E,y}^{(0)}\cos^2\alpha}{1-I^{(0)}_{yy}}
 \right]
\nonumber\\
&=&\nu \mu_{\rm B}\,B_z\,\left(\frac{m_e}{m^*}\right)
\frac{2\Omega\tau}{1-I^{(0)}_{zz}} \times\nonumber\\&\times&
\left[c^{(1)}_{E,z}(\alpha)
+\frac{ i^{(1)}_{zx}c_{E,x}^{(0)}\sin^2\alpha}{1-I^{(0)}_{xx}}
+\frac{ i^{(1)}_{zy}c_{E,y}^{(0)}\cos^2\alpha}{1-I^{(0)}_{yy}}
 \right]
 \ ,\nonumber\\
\end{eqnarray} 
where $\mu_{\rm
B}\equiv e/(2m_e c)$ and $m_e$ is the bare electron mass.
The last form is most convenient for dimensional analysis and plotting, as the factor
in square brackets is dimensionless.   

Finally, for the Zeeman term we obtain
\begin{eqnarray}\label{eq:s_z_Zeeman}
s_z^{(1),\rm Zeeman} &=& \frac{\tau}{1-I^{(0)}_{zz}} \, \frac{\nu\,g\,\mu_{\rm B}}{2\tau}
\left[1- a I^{(0)}_{zz} \right] B_z\nonumber\\
&=&\frac{1- a I^{(0)}_{zz} }{1-I^{(0)}_{zz}} \cdot \frac{\nu\,g\,\mu_{\rm B}}{2}
 B_z\ .
\end{eqnarray} 

{\it Discussion.} 
We observe that both parts of the out-of-plane spin polarization $s_z$, i.e., the orbital part 
given by Eq.~(\ref{eq:s_z_orbital}) and the Zeeman part given by Eq.~(\ref{eq:s_z_Zeeman})
can be regarded as linear response to the out-of-plane magnetic field $B_z$. 
 Thus our analysis amounts to a calculation of the susceptibility $\chi(\omega,\vecbf q)$ so that 
$s_z=\chi B_z$.  

\begin{figure}[tbh]
\center{\includegraphics[width=0.7\columnwidth]{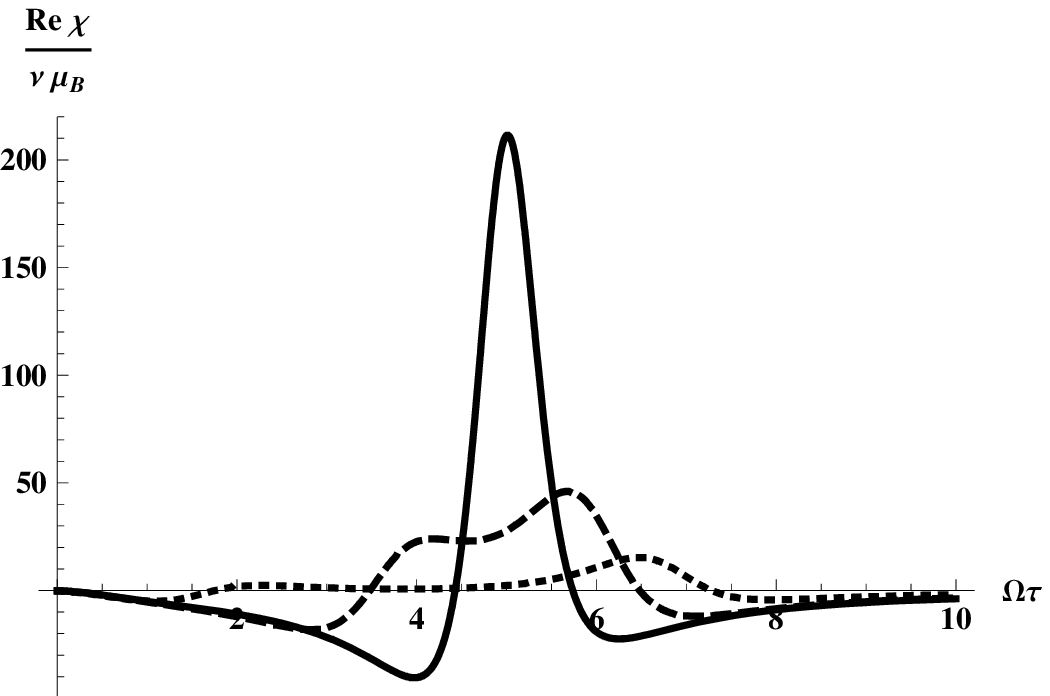}
\includegraphics[width=0.7\columnwidth]{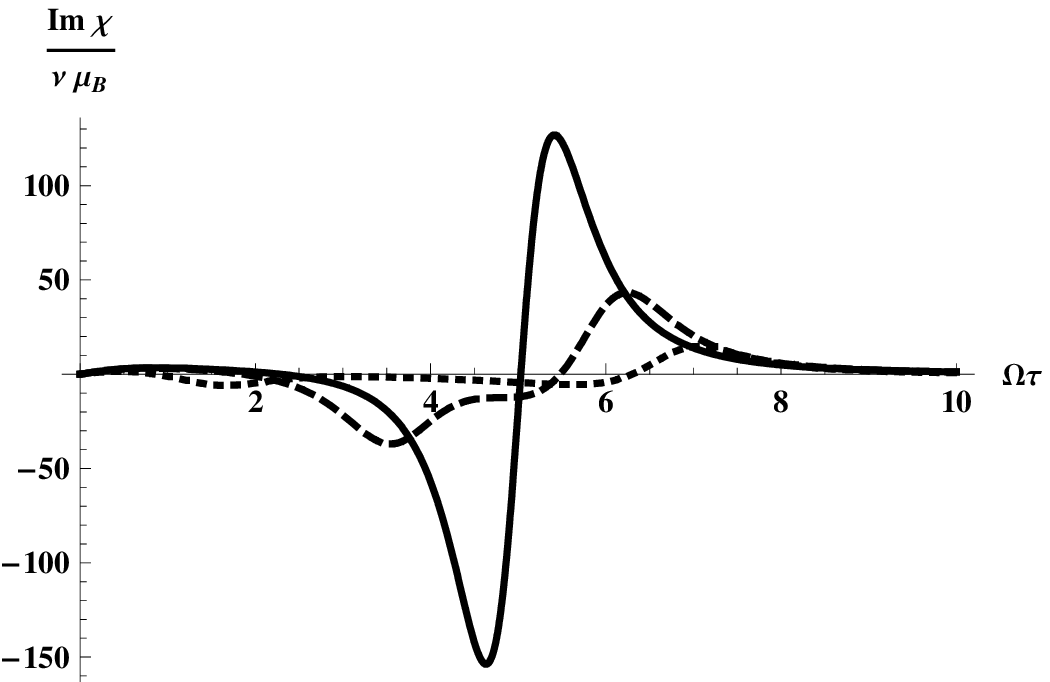}}
\caption[]{\label{fig:plot_averaged} Real and imaginary parts of the total 
(orbital plus Zeeman) spin susceptibility $\chi\equiv s_z/B_z$ for 
$2\Delta_F\tau = 5$. Solid lines: $T_\alpha=0.1$, dashed lines: $T_\alpha=0.5$, 
dotted lines: $T_\alpha=0.9$.
Parameters
assumed as in GaAs: $m_e/m^* \approx 15$, $g=-0.44$. 
The results are plotted for $\alpha = \pi/4$ which also corresponds to the averaged over
$\alpha$ susceptibility.}
\end{figure}

\begin{figure}[tbh]
\center{\includegraphics[width=0.7\columnwidth]{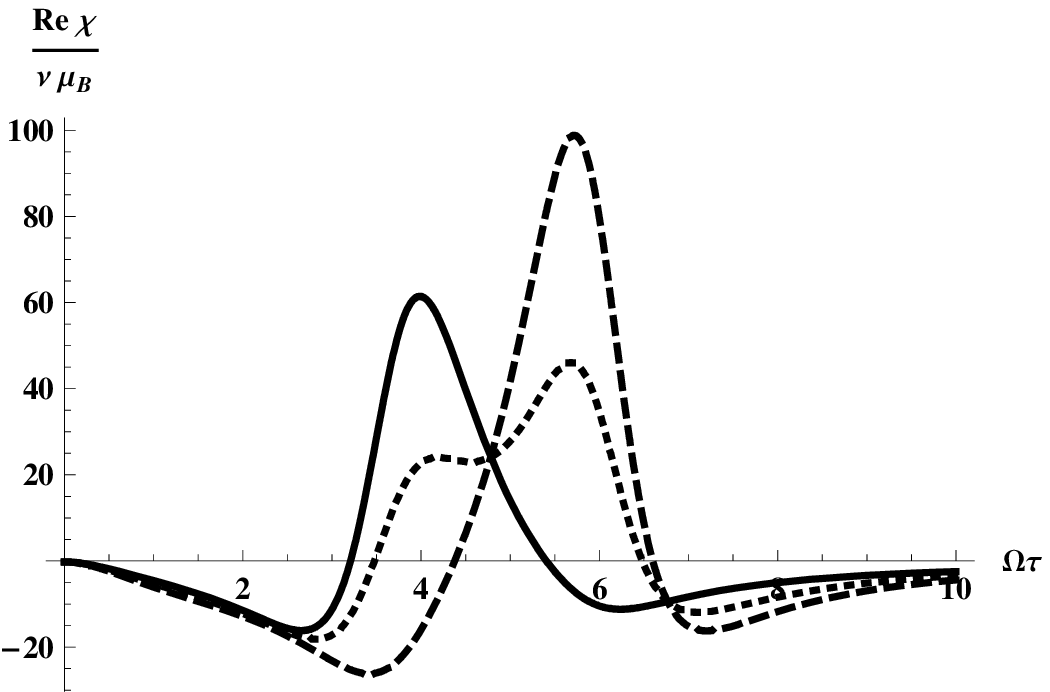}
\includegraphics[width=0.7\columnwidth]{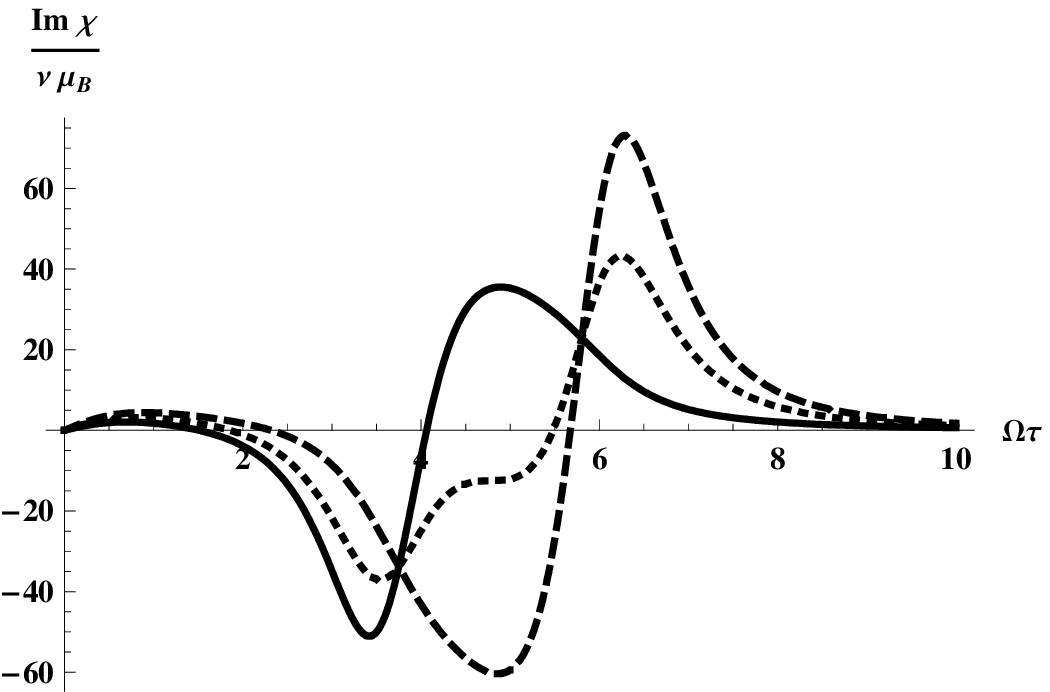}}
\caption[]{\label{fig:plot_angles} Real and imaginary parts of the total 
(orbital plus Zeeman) spin susceptibility $\chi\equiv s_z/B_z$ for 
$2\Delta_F\tau = 5$ and $T_\alpha=0.5$. Solid lines: $\alpha=0$, 
dashed lines: $\alpha=\pi/2$, dotted lines: $\alpha=\pi/4$. Parameters
assumed as in GaAs: $m_e/m^* \approx 15$, $g=-0.44$. 
}
\end{figure}

We observe that in the clean limit, i.e., for $\Delta_F > \tau^{-1}$, the orbital 
susceptibility greatly dominates over the Zeeman one. As one can see in 
Figs.~\ref{fig:plot_averaged} and \ref{fig:plot_angles}, for experimentally relevant 
parameters, the susceptibility $\chi$ exceeds the Pauli susceptibility by a factor
of order hundreds. 

In the vicinity of  $T_\alpha = 0$ we reproduce the results of Ref.~\cite{ShnirmanMartin} and the susceptibility 
is peaked around $\Omega = 2\Delta_F$. For $T_\alpha$ substantially different from zero a double peak structure 
develops with the positions $\Omega = 2\sqrt{1\pm T_\alpha}\Delta_F$. Here 
the pole singularity present at $T_\alpha = 0$ splits into two square-root singularities corresponding to 
zeros of function $R(\Omega)$ at $\Omega = 2\sqrt{1\pm T_\alpha}\Delta_F - i/\tau$.  
One or the other peak are emphasized depending on the 
angle $\alpha$ as seen in Fig.~\ref{fig:plot_angles}.

We obtained our results for a microwave with a given direction of the wave-vector 
$\vecbf q$, i.e., for a given angle $\alpha$. The most obvious way to observe 
the orbital contribution to the spin susceptibility would be by applying a 
homogeneous oscillating magnetic field $B_z$, e.g., by putting the 
sample into a magnetic coil. Such a field corresponds to an equal superposition 
of plane waves with all possible 
wave vectors $\vecbf q$ laying in the $xy$ plane. To obtain the orbital spin 
response in this case one should just average over $\alpha$, i.e., substitute 
$\langle \cos 2\alpha \rangle = 0$ and   $\langle \cos^2 \alpha \rangle = 
\langle \sin^2 \alpha \rangle=1/2$. This is what we plot in 
Fig.~\ref{fig:plot_averaged}. Note, that for $T_\alpha=0$, i.e., for pure Rashba or Dresselhaus coupling, the response is 
$\alpha$-independent and the averaging brings nothing new. On the other hand, 
when the two couplings are of comparable strength, 
the susceptibility strongly depends on $\alpha$ (see Fig.~\ref{fig:plot_angles}) and 
averaging over $\alpha$ 
can change the result considerably. At $T_\alpha=1$ the orbital susceptibility vanishes. 

We thank I. Martin for numerous encouraging discussions. 
 
\bibliography{ref}

\end{document}